\documentclass[a4paper,twocolumn,10pt,accepted=2026-06-12]{quantumarticle}
\pdfoutput=1
\usepackage[utf8]{inputenc}
\usepackage[english]{babel}
\usepackage[normalem]{ulem}
\usepackage[T1]{fontenc}
\usepackage{amsmath}
\usepackage{amssymb}
\usepackage{amsthm}
\usepackage{hyperref}
\hypersetup{
  citecolor=magenta,
}
\usepackage{physics}
\usepackage[numbers,sort&compress]{natbib}

\usepackage{tikz}
\usetikzlibrary{positioning}
\usetikzlibrary{calc}
\usetikzlibrary{fit}
\usetikzlibrary{arrows.meta}

\usepackage{graphicx}

\DeclareMathOperator*{\argmax}{arg\,max}

\usepackage{soul}

\begin{document}

\title{Machine Learning Decoding of Circuit-Level Noise for Bivariate Bicycle Codes}

\author{John Blue}
\affiliation{Department of Physics, Massachusetts Institute of Technology}
\affiliation{The NSF AI Institute for Artificial Intelligence and Fundamental Interactions}
\author{Harshil Avlani}
\affiliation{Department of Physics, Massachusetts Institute of Technology}
\author{Zhiyang He}
\affiliation{Department of Mathematics, Massachusetts Institute of Technology}
\author{Liu Ziyin}
\affiliation{Department of Physics, Massachusetts Institute of Technology}
\affiliation{Physics \& Informatics Laboratories, NTT Research}
\author{Isaac L. Chuang}
\affiliation{Department of Physics, Massachusetts Institute of Technology}

\maketitle
\begin{abstract}

Fault-tolerant quantum computers will depend crucially on the performance of the classical decoding algorithm which takes in the results of measurements and outputs corrections to the errors inferred to have occurred.
Machine learning models have shown great promise as decoders for the surface code; however, this promise has not yet been substantiated for the more challenging task of decoding quantum low-density parity-check (QLDPC) codes.
In this paper, we present a recurrent, transformer-based neural network designed to decode circuit-level noise on Bivariate Bicycle (BB) codes.
For the \([[72,12,6]]\) BB code, at a physical error rate of \(p=0.1\%\), our model achieves logical error rates almost \(5\) times lower than belief propagation with ordered statistics decoding (BP-OSD), and roughly \(5\) times larger than a most-likely error decoder.
Moreover, while BP-OSD has a wide distribution of runtimes with significant outliers, our model has a consistent runtime and is an order-of-magnitude faster than the worst-case times from a benchmark BP-OSD implementation.
On the $[[144,12,12]]$ BB code, our model obtains worse logical error rates but maintains the speed advantage.
These results provide initial evidence that machine learning decoders can out-perform conventional decoders on small QLDPC codes, but suggest more complex architectures and/or training procedures are necessary to scale to larger code sizes.
\end{abstract}

\section{Main}
\subsection{Introduction}
\label{sec:intro}

Quantum error correction (QEC) is expected to be a necessary component of large-scale quantum computers. 
The standard procedure of QEC is to perform a series of measurements (in a process often referred to as \textit{syndrome extraction}), producing a set of results called a \textit{syndrome}. A classical \textit{decoder} then takes in the syndrome and makes a prediction as to what error occurred.
During the operation of a quantum computer, the decoder must be able to output corrections as fast as the measurements are performed (so-called \textit{real-time decoding}).
Otherwise, the backlog of measurement results will grow, resulting in a slowdown of the computation that is exponential in the number of non-Clifford gates in the circuit~\cite{terhal2015error}.
In recent years, significant work has started to tackle the challenge of real-time decoding specifically for the case of the surface code~\cite{das2022accurate,skoric2023parallel,bombin2023modular,tan2023scalable,wu2023fusion,zhang2023scalable,liyanage2023scalable,caune2024demonstrating,higgott2025sparse,barber2025real,wu2025micro}.

Meanwhile, fault-tolerance is being improved through the construction of quantum low-density parity-check (QLDPC) codes~\cite{tillich2013quantum,Leverrier2015,hastings2021fiber,panteleev2021quantum,Breuckmann2021balanced,breuckmann2021quantum,panteleev2021degenerate,panteleev2022asymptotically,leverrier2022quantum,dinur2023good,xu2024constant,bravyi2024highthreshold}.
Compared to the surface code, which encodes one logical qubit per block, QLDPC codes offer higher encoding rates, making them promising candidates to greatly reduce the space overhead of large-scale fault-tolerant quantum computation. 
The leading decoder for small-to-medium scale QLDPC codes\footnote{Linear-time decoders have been developed for asymptotic constructions of QLDPC codes ~\cite{leverrier2015quantum,Fawzi2020,Gu2023,Leverrier2023,dinur2023good,Gu2024}. 
However, they are not yet competitive in practical regimes.} is the Belief Propagation with Ordered Statistics Decoding (BP-OSD) algorithm proposed by~\cite{panteleev2021degenerate}.
While BP-OSD can be applied to any QLDPC code and reliably produces low logical error rates~\cite{roffe2020decoding}, the OSD step has a worst case runtime cubic in the size of the code. 
Consequently, in practice the runtime of BP-OSD has high variance and significant outliers.
Several works have investigated faster alternatives to OSD ~\cite{iolius2024almost,iolius2024closed,hillmann2024localized,gong2024low,wolanski2025ambiguity,ott2025decision,beni2025tesseract}.
However, achieving the speed and logical error rates needed for real-time decoding remains a challenging open problem for QLDPC codes.

Machine learning (ML) has emerged as a promising alternate method for decoding topological codes such as the surface code~\cite{torlai2017neural,varsamopoulos2017decoding,krastanov2017deep,chamberland2018deep,baireuther2018machine,baireuther2019neural,andreasson2019error,maskara2019advantages,sweke2020reinforcement,ni2020neural,varsamopoulos2020decoding,varsamopoulos2020comparing,colomer2020reinforcement,fitzek2020deep,wagner2020symmetries,overwater2022neural,ueno2022neoqec,meinerz2022scalable,wang2023transformerqec,cao2023qecgpt,chamberland2023techniques,gicev2023scalable,lange2023data,bausch2024learning,varbanov2025neural}.
ML has desirable features for real-time decoding, such as a constant runtime for a given code size, unlike standard approaches which can suffer from syndrome-dependent runtimes.  ML decoders can also take advantage of rapidly improving and widely available commodity hardware accelerators~\cite{reuther2022accelerator,boutros2024field}.

However, training an ML decoder for larger distance codes is well known to be challenging~\cite{sweke2020reinforcement,varsamopoulos2020comparing,varbanov2025neural,lange2023data,bausch2024learning}.
As the size of the code grows, the number of possible syndromes grows exponentially, and thus, in order for the model to see a representative sample of possible syndromes, the amount of training data must also grow exponentially~\cite{gruber2017deep,varsamopoulos2020comparing}.
This issue is exacerbated when moving from simple code capacity noise (in which syndrome measurements are assumed to be perfect) to more complicated circuit-level noise (which includes the possibility of faulty measurements and error propagation),
as the size of the input space also grows exponentially with the number of syndrome measurement rounds.
Furthermore, the relationship between errors and syndromes becomes more complicated. In the surface code, models that have performed well on larger distance codes have taken advantage of the local nature of the stabilizer operators, either through the use of convolutional layers~\cite{zhang2023scalable,bausch2024learning}, or by using the ML model as a ``local decoder'' that is then followed by a non-ML ``global decoder''~\cite{meinerz2022scalable,chamberland2023techniques}.
This local structure is lost when moving to QLDPC codes.
Additionally, for an ML model that attempts to approximate a maximum likelihood decoder, \textit{e.g.} by predicting logical errors, the output space has a size exponential in the number of logical qubits, and thus a QLDPC code will require a much larger number of output classes (\emph{i.e.} each of the \(2^k\) possible logical errors in a memory experiment) than a surface code.

In this work, we provide an initial data point in space of generative model architectures for decoding circuit-level noise on QLDPC codes by training a recurrent, transformer-based ML decoder for the bivariate bicycle (BB) codes introduced in~\cite{bravyi2024highthreshold}. Specifically, we focus on addressing circuit-level noise through introduction of three advances to ML QLDPC decoding:
\begin{itemize}
\item extending to quantum codes the use of code-aware self-attention, originally proposed for classical codes~\cite{choukroun2022error}, to accelerate ML decoder training and increase decoder robustness;

\item employing a recurrent neural network architecture~\cite{bausch2024learning} to improve trainability through reduction of the input size per model iteration; and 

\item choosing a conditional probability distribution as the model output, following~\cite{cao2023qecgpt}, to allow the ML decoder to (ideally) approximate a maximum likelihood decoder without requiring an exponentially large number of output classes, via autoregressive prediction of whether or not errors that anti-commute with the measured logical operators have occurred in the circuit.
\end{itemize}

Comparing our ML decoder with a specific benchmark BP-OSD code for circuit-level noise decoding, on the \([[72,12,6]]\) BB code, our model outperforms the benchmark both in terms of logical error rate and worst-case runtime, with the caveat that BP-OSD was run on a CPU, and the ML decoder was run on a GPU. There is also still a large gap between the ML decoder and a most-likely error decoder on this code, suggesting additional room for improvement for the ML decoder. On the \([[144,12,12]]\) code, our model achieves similar logical error rates to the benchmark for physical error rates slightly below psuedo-threshold, although BP-OSD begins to obtain much better logical error rates further below psuedo-threshold. Our results serve as an initial baseline for future ML decoders on QLDPC codes to compare against, and provide evidence that ML decoders can be useful on small QLDPC codes; however it appears that further architecture and/or training improvements are required in order to obtain desired logical error rates on larger codes.

These results are presented below, beginning with a review of prior work in ML decoders (Section~\ref{subsec:prior}), followed by the comparison scenario and model performance results (Section~\ref{sec:model-performance}) and an outlook discussion (Section~\ref{subsec:discussion}).  Details about the methods employed, including background on stabilizer codes, transformers, and the specification of the decoding task, are presented in Section~\ref{sec:methods}, together with our model architecture, the code-aware self-attention mechanism, our training methodology, and our experimental procedure.
We include further details about the model architecture in appendix~\ref{sec:model-arch-deta}, and specific model and training hyperparemeters in appendix~\ref{sec:training-details}.

\subsection{Prior Work}
\label{subsec:prior}

The literature on ML decoding of quantum codes is broad, with many different models attempting to decode a variety of different noise models~\cite{torlai2017neural,varsamopoulos2017decoding,krastanov2017deep,chamberland2018deep,baireuther2018machine,baireuther2019neural,andreasson2019error,maskara2019advantages,sweke2020reinforcement,ni2020neural,varsamopoulos2020decoding,varsamopoulos2020comparing,colomer2020reinforcement,fitzek2020deep,wagner2020symmetries,overwater2022neural,ueno2022neoqec,meinerz2022scalable,wang2023transformerqec,cao2023qecgpt,chamberland2023techniques,gicev2023scalable,lange2023data,bausch2024learning,varbanov2025neural}.
Several approaches for ML decoding of surface codes are most relevant to our work.
In~\cite{cao2023qecgpt}, the authors developed an autoregressive transformer model for decoding code capacity noise on surface codes, which outperformed the minimum weight perfect matching (MWPM) decoder~\cite{edmonds1965paths,higgott2025sparse} for distances three, five, and seven.
Notably, the model in~\cite{cao2023qecgpt} continued to outperform MWPM when introducing defects into the surface and thereby increasing the number of logical qubits.

However, possibly due to the challenges raised above,~\cite{cao2023qecgpt} did not extend their results to circuit-level noise or codes beyond the surface code.
In~\cite{bausch2024learning}, the authors used a recurrent, transformer-based model to decode circuit-level (and experimental) noise on the surface code, on distances up to eleven and found the model produced logical error rates better than a correlated MWPM decoder.

Some initial works have begun to explore ML decoders for QLDPC codes.
All three of~\cite{maan2024machine}, \cite{ninkovic2024decoding}, and \cite{gong2024graph} investigated using graph neural networks to decode code capacity noise for various families of QLDPC codes, and~\cite{liu2019neural} used neural belief-propagation.
However, again likely due to the practical ML model training challenges described above, none of these works have extended their results to circuit-level noise.

\begin{figure*}[ht]
  \centering
  \includegraphics{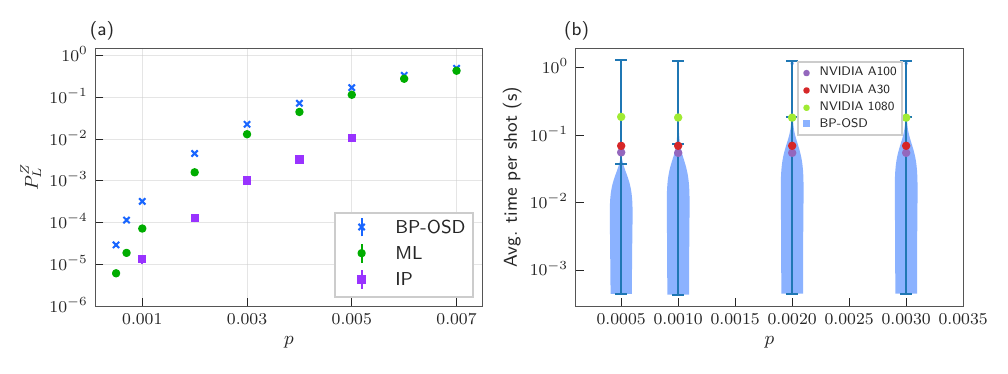}
  \caption{A comparison of the ML model and BP-OSD on
    the \([[72,12,6]]\) BB code.
    The exact hyper parameters and training details of the model shown can be found in Appendix~\ref{sec:training-details}.
    (a) The logical \(Z\) error rate of the memory experiment vs physical error rate for both the ML decoder and BP-OSD, as well as the integer programming  (IP) decoder from \cite{beni2025tesseract}, which is a most-likely error decoder. 
    We note that the error rate is for the entire memory experiment, and not scaled per-round.
    Third order OSD was used when generating this data.
    (b) A comparision of times across physical error rates for the ML decoder and BP-OSD. The violin plot shows the distribution of times of the BP-OSD decoder for \(50,000\) shots at each error rate, measured on an AMD EPYC 9654 96-Core Processor (2.4 GHz).
    Zero-order OSD was used to generate this data.
    The points show the average ML time to decode 1,000 shots at each error rate, measured on the indicated GPU, with a batch size of one.}
  \label{fig:72-lers-times}
\end{figure*}
\subsection{Model Performance}
\label{sec:model-performance}

In this section, we describe the results of our evaluations of the ML decoder.
We begin with a brief description of our chosen codes, comparison with BP-OSD, and the decoding task performed, before analyzing the results on the \([[72,12,6]]\) and \([[144,12,12]]\) BB codes.

We decode two instances of (BB) codes; a \([[72,12,6]]\) code and a \([[144,12,12]]\) code.
These codes have much better encoding rates than the surface code, similar pseudo-thresholds (roughly \(0.7\%\)), and in the case of the \([[144,12,12]]\) code, a method of performing the full logical Clifford group through Pauli measurements~\cite{cross2024improved}, making them promising candidates for a near-term fault-tolerant quantum processor.

In order to determine how well the model is performing, we compare it with an open source BP-OSD implementation from Roffe et al~\cite{roffe2020decoding,roffe2022ldpc}.
As described in the introduction, BP-OSD is the standard decoding algorithm used to determine the performance of QLDPC codes (\textit{e.g.}~\cite{xu2024constant,bravyi2024highthreshold}), and this particular implementation is often used as a comparison with new decoders~\cite{iolius2024almost,iolius2024closed,wolanski2025ambiguity,ott2025decision,beni2025tesseract}.  We recognize that this BP-OSD implementation may not be the fastest possible, but its accessibility, history, and generally good performance make it an apt baseline benchmark to choose for comparison. 

We compare our model and BP-OSD on memory experiments -- repeated rounds of syndrome measurement using a standard circuit-level depolarizing noise model.
For each code, we evaluate both decoders on a memory experiment in which the number of noisy syndrome measurement rounds is equal to the distance.
For the \([[72,12,6]]\) code we also investigate memory experiments with larger numbers of noisy syndrome measurement rounds.
More information on the exact structure of the circuits and the noise model can be found in Section~\ref{sec:memory-experiment}.
We note that, following~\cite{bravyi2024highthreshold}, we use BP-OSD to decode syndromes only from the \(X\) check operators, whereas the ML model uses syndromes from both the \(X\) and \(Z\) check operators. 
\footnote{While BP-OSD can in principle be used to decode syndromes from both \(X\) and \(Z\) check simultaneously, a straightforward implementation has impractically slow runtimes.
In our numerical simulations, decoding 23,000 shots of a memory experiment with \(12\) noisy rounds of syndrome measurement with the \([[144,12,12]]\) BB code at a physical error rate of \(p=0.1\%\) had an average runtime of \(68.3\)s, and a maximum of runtime of \(468\)s. The simulations were run on an AMD EPYC 9654 96-Core Processor (2.4 GHz).}

For each memory experiment, we compare the ML decoder and BP-OSD on two metrics: the logical error rate and the runtime.
The logical error rate is the fraction of the time that the decoders correctly predict whether or not errors that anti-commute with the final logical measurements at the end of the memory experiment have occurred.
The runtime is the time it takes for the decoder to make these predictions.
As we describe in more detail below, interpreting a comparison of the runtimes is challenging, because the decoders are run on different kinds of processors (BP-OSD on CPU, and ML on GPU) and employ different programming libraries and language components. Additionally, we note that when evaluating BP-OSD on logical error rate, we used third order OSD, while when evaluating the runtime of BP-OSD, we used zero-order OSD.

Figure~\ref{fig:72-lers-times} shows a comparison of the machine learning model and BP-OSD on the \([[72,12,6]]\) code in a memory experiment with six noisy rounds of syndrome measurement.
We also include the results of an integer programming decoder from \cite{beni2025tesseract} (using syndromes from both the \(X\) and \(Z\) checks), which is a most-likely error decoder. While the runtime of such a decoder makes it impractical for most use cases, it allows us to determine how close to optimal the ML decoder is for this task.
As shown in part (a) of the figure, at all tested physical error rates, the ML decoder has a lower logical error rate than BP-OSD-3, and this difference grows as the physical error rate gets smaller.
We note that our model was only trained on data generated at a physical error rate of \(0.6\%\), and was able to generalize well to lower error rates. For \(p=0.1\%\), the ML decoder has a logical error rate roughly \(4.5\) times lower than BP-OSD-3, and is \(5.4\) times larger than the most-likely error decoder.

A comparison of runtimes is shown in Figure~\ref{fig:72-lers-times}(b). We see that the ML decoder has a  runtime similar to the average runtime of BP-OSD-0, and roughly an order of magnitude faster than the worse-case instances for BP-OSD-0.
However, we must caveat this discussion of runtimes with the disclaimer that comparisons of software implementations of decoding algorithms likely would not reflect the eventual performance of hardware decoders.
The times shown in Figure~\ref{fig:72-lers-times}(b) were obtained by running BP-OSD-0 on a CPU, and the ML model on several different GPUs.
An actual decoder running concurrently with a quantum computer will likely be implemented on specialized hardware (such as a field programmable gate array or application specific integrated circuit) and achieve runtimes much faster than those shown in Figure~\ref{fig:72-lers-times}(b). 
However, Figure~\ref{fig:72-lers-times}(b) does demonstrate one feature of BP-OSD that is likely to persist in hardware implementations; the range of runtimes varies drastically, with particularly slow instances (occurring when BP fails to converge and OSD is used) taking up to an order of magnitude longer than an average instance.
ML inference times, on the other hand, are consistent across different inputs.

\begin{figure}[ht]
  \centering
  \includegraphics{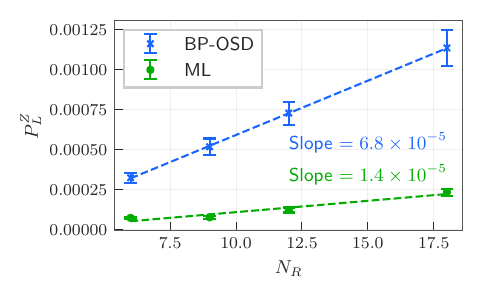}
  \caption{A comparison of BP-OSD-3 and ML decoders over a range of \(N_R\) values.
  Each plotted ML point represents the logical error rate of a model trained on a memory experiment with that number of noisy rounds.
  Error bars indicate one standard deviation (\(\sqrt{p(1-p)/n}\) for \(p=k/n\) with \(k\) observed logical errors and \(n\) shots).}
  \label{fig:72-12-6-multi}
\end{figure}

We also investigate the performance of the model for longer memory experiments.
\footnote{While only \(d\) rounds of syndrome measurement are required for error correction to be fault-tolerant, in practice, one might like to keep a logical qubit coherent for longer that \(t_{SM}\cdot d\), where \(t_{SM}\) is the time for a single round of syndrome measurement. In such a scenario one would have to decode (potentially significantly) more than \(d\) rounds of syndromes.}
Figure~\ref{fig:72-12-6-multi} compares the logical error rate obtained by BP-OSD-3 and the ML decoder for a number of noisy syndrome measurement rounds (\(N_R\)) ranging from six to \(18\), using a physical error rate of \(p=0.1\%\).

We note that a given ML model performed poorly for values of \(N_R\) much different than the value it was trained at (\textit{e.g.} our model trained for \(N_R=9\) had a higher logical error rate than BP-OSD-3 for \(N_R=18\)).
In Figure~\ref{fig:72-12-6-multi}, we show the results for separate ML models
trained at each plotted \(N_R\) value.
From these graphs, we can perform a linear fit to get an estimate of the ``logical error rate per round'' for both the ML model and BP-OSD-3.
From this fit, we see that the ML decoder offers greatly improved logical error suppression as the number of syndrome measurement rounds grows. 
Effectively training a model that is able to obtain good logical error rates for \(N_R > 18\), as well as a single model that can perform well over a range of \(N_R\) are important tasks for future work.

\begin{figure*}[ht]
  \centering
  \includegraphics{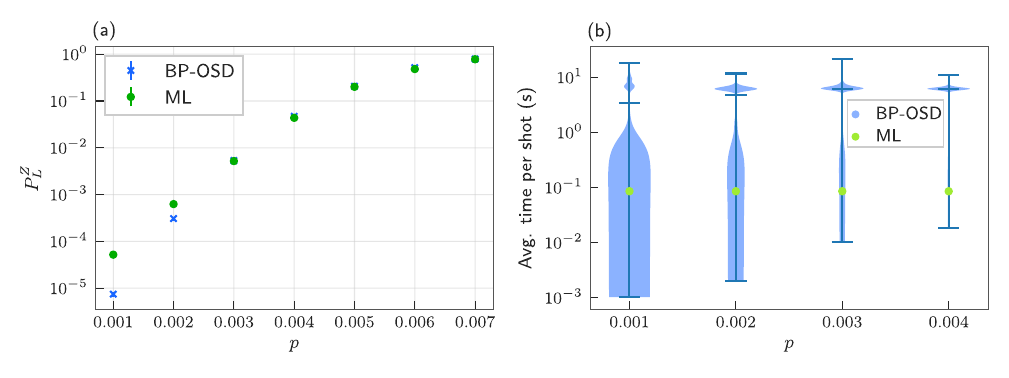}
  \caption{A comparison of the ML model and BP-OSD on the \([[144,12,12]]\) BB code.
  The exact hyper parameters and training details of the model shown can be found in appendix~\ref{sec:training-details}. (a) The logical \(Z\) error rate of the memory experiment vs physical error rate for both the ML decoder and BP-OSD. 
  We note that the error rate is for the entire memory experiment, and not scaled per-round.
  Third order OSD was used when generating this data.
  (b) A comparison of times across physical error rates for the ML decoder and BP-OSD.
  The violin plots show the distribution times of the BP-OSD decoder for 50,000 shots at each error rate, measured on a AMD EPYC 9654 96-Core Processor (2.4 GHz).
  Zero-order OSD was used to generate this data.
  The points show the average ML time to decode 1,000 shots at each error rate, measured on an NVIDIA A100 GPU with a batch size of one.}
  \label{fig:144-results}
\end{figure*}

Initial results of the model on the \([[144,12,12]]\) BB code are shown in Figure~\ref{fig:144-results}.
For this model, we primarily trained on data generated at a physical error rate of \(p=0.6\%\).
However, we found that performance at lower physical error rates was improved by training at \(p=0.4\%\) as well (a full description of the training hyperparameters used for this model can be found in Appendix~\ref{sec:training-details}).
As shown in the figure, the ML model obtains similar logical error rates to BP-OSD-3 for physical error rates between \(p=0.3\%\) to \(p=0.7\%\), and it begins to perform worse for \(p < 0.3\%\).
At \(p = 0.1\%\), the logical error rate is roughly seven times larger than BP-OSD-3.
This seems to suggest that the ML decoder is not correcting up to same distance as BP-OSD-3.
As shown in part (b) of the figure, for a physical error rate of \(p=0.1\%\), the runtime of the ML decoder is over an order of magnitude smaller than the mean runtime of BP-OSD-0.
We comment on possible changes to the model to eliminate the difference in logical error rate in the next section.

\subsection{Discussion and Outlook}
\label{subsec:discussion}

In this work, we have demonstrated a transformer-based machine learning model designed specifically to tackle some of the challenges in decoding QLDPC codes. On the \([[72,12,6]]\) code, the model outperforms BP-OSD both in terms of logical error rate and worst-case runtime, demonstrating the utility of the techniques introduced.

There are a wide range of further avenues to explore.
One immediate route is improving the performance of the ML model on the \([[144,12,12]]\) BB code at lower physical error rates.
Our results on the \([[144,12,12]]\) code indicate that our chosen architecture has trouble scaling to larger code sizes, suggesting some combinations of architecture improvements and increased model/training scale. Indeed, during the writing of this paper, the authors were made aware of the results of \cite{hu2025efficient}, which employed, among other changes, more modern linear attention mechanisms and almost \(7\times\) more training data
\footnote{The number of training examples for the model in \cite{hu2025efficient} was obtained by multiplying the number of pre-training steps (\(10^7\)) by the batch size (\(256\)), both listed in Table S2, yielding \(2.56 \times 10^9\). For the model in this paper, we multiply the number of examples per epoch \((16,384)\) by the total number of epochs (\(23,000\) as shown in Table \ref{tab:4}), to get roughly \(3.8 \times 10^8\).}, 
to obtain better accuracy than BP-OSD on the \([[144,12,12]]\) code. Additionally, after the original posting of this paper, \cite{liu2025decodingquantumlowdensity} also obtained better logical error rates than BP-OSD using a diffusion based model (as well as a multi-stage training procedure similar to the one described in Section \ref{sec:training}). While our work does not achieve as good logical performance as that of \cite{hu2025efficient} and \cite{liu2025decodingquantumlowdensity}, we believe it complements them in beginning to explore the space of generative model architectures and understanding the requirements for a model and training scheme that can obtain low logical error rates on larger codes.

As another direction, it would be interesting to further experiment with the physical error rate used to generate the training data for the model.
As described in Section~\ref{sec:model-performance}, we found it sufficient to train at a physical error rate of \(p=0.6\%\) for the \([[72,12,6]]\) code, but useful to train at a lower error rate of \(p=0.4\%\) for the \([[144,12,12]]\) model.
In our experiments, we found no additional benefit to training at even lower error rates.
We hypothesize this is because the training data at low physical error rates is very sparse - in a given batch, a well-trained model will make very few errors and not obtain a useful gradient.
Perhaps doing some sort of specialized sampling (\textit{e.g.} as in~\cite{bravyi2013simulation}) could help alleviate this issue and make training at lower physical error rates feasible.

Looking further into the future, an important next step for machine learning decoders will be demonstrations of decoding logical computations on QLDPC codes, as opposed to just memory. 
As we describe in Section~\ref{sec:decoding-task}, an ML decoder that can predict logical measurement flips is sufficient for Pauli-based computation, which has emerged as a practically promising computational model for QLDPC codes through the extractor architectures~\cite{he2025extractors}. 
However, in order for this decoder to work in real-time, it must be able to produce some logical measurement flips before the computation is done.
Furthermore, in the setup we have described, we assume the model is trained on the results of every syndrome measurement in the computation.
For a large computation involving millions of operations, this is impractical.
Instead, we will need machine learning models capable of predicting logical measurement flips when given access to only a
small subset of the measurement outcomes.
This is similar in spirit to the problem solved by parallel window decoding~\cite{skoric2023parallel,bombin2023modular,tan2023scalable} for the surface code, and we can imagine a similar set up for our ML model
decoding QLDPC codes, where different models are tasked with decoding specific regions of the space-time Tanner graph, and the different models pass some representation of the measurement data in their region (such as the output of the encoder block in our model) to the models decoding adjacent regions. The code-aware self-attention mechanism which we employ could likely be generalized to realize a kind of computation-aware attention mechanism, for these purposes.  We leave a detailed implementation of these ideas for future work.

\section{Methods}
\label{sec:methods}
In this section we provide additional details on our work that were omitted in the main text.
We begin with a brief background on the aspects of QEC (Section~\ref{sec:qec}) and transformers (Section~\ref{sec:transformers}) needed for this work.
We then describe in more depth the decoding task (Section~\ref{sec:decoding-task}), before moving into how we solve that decoding task with our chosen model architecture (Section~\ref{sec:model-architecture}), training procedure (Section~\ref{sec:training}), and code-aware self-attention (Section~\ref{sec:code-aware-self}). 
Finally, we include the details of our memory experiments in Section~\ref{sec:memory-experiment}.

\subsection{Quantum Error Correction}
\label{sec:qec}

Here we give a brief overview of QEC, describing a common class of quantum error correcting codes that includes BB codes (stabilizer codes) and two common noise models often used when evaluating codes.
For a full overview of QEC, we recommend~\cite{gottesman2016surviving}.

We use the following notation to denote the Pauli matrices,
\begin{align*}
  I &= \begin{pmatrix} 1 & 0 \\ 0 & 1 \end{pmatrix}, \quad
       X = \begin{pmatrix} 0 & 1 \\ 1 & 0 \end{pmatrix} \\
  Y &= \begin{pmatrix} 0 & -i \\ i & 0 \end{pmatrix}, \quad
       Z = \begin{pmatrix} 1 & 0 \\ 0 & -1 \end{pmatrix}.
\end{align*}
Let \(\mathcal{P}_n = \langle i, X_1, X_2, \dots, X_n, Z_1, Z_2, \dots, Z_n \rangle\) denote the Pauli group on \(n\) qubits, where, \textit{e.g.} \(X_2\) indicates an operator that applies the Pauli \(X\) operator to qubit \(2\), and acts as the identity on all other qubits. The \emph{weight} of a Pauli operator is the number of qubits on which it acts non-trivially (so \(X_2\) has weight one, and \(X_2X_{10}\) has weigh two).
A stabilizer code~\cite{gottesman1997stabilizer} is described by an Abelian subgroup \(\mathcal{S}\) of the Pauli group, such that \(-I \notin \mathcal{S}\). 
The code \(\mathcal{C}\) is the space of shared \(+1\) eigenvectors of the operators in \(\mathcal{S}\).

Let \(\left\{ M_1, \dots, M_s \right\}\) be a subset of \(\mathcal{S}\) that generates the entire group.
Errors are detected by measuring this set of stabilizer elements (which are also called ``check operators'').
This set of measurements is called a syndrome.
For example, let \(\ket{\psi} \in \mathcal{C}\) be a codeword. 
If a Pauli error \(E\) occurs such that \(M_j E = -E M_j\) for a stabilizer element \(M_j\), then \(E \ket{\psi}\) is now a \(-1\) eigenvector of \(M_j\), and measuring \(M_j\) indicates the presence of an error.
Furthermore, it can be shown that for a code on \(n\) qubits, the number of logical qubits encoded by a stabilizer code \(\mathcal{S}\) with \(s\) independent generators is given by \(k = n - s\).

The set of errors that are undetectable by a given stabilizer code is given by the set of Pauli operators that commute with every element of the stabilizer, which in the case of the Pauli group can be shown to be equivalent to the normalizer \(N(\mathcal{S})\).
However, elements of the stabilizer group do not affect codewords, and thus the errors that can
actually affect the codespace are given by \(N(\mathcal{S}) \setminus \mathcal{S}\).
This is the set of logical Pauli operators of the code, as they perform nontrivial operations on the codespace, and furthermore \(N(\mathcal{S}) / \mathcal{S} \cong \mathcal{P}_k\).
The distance of a stabilizer code is given by the weight of the smallest weight logical operator.
It is standard notation to use \([[n, k, d]]\) to describe a stabilizer code with \(n\) physical qubits, \(k\) logical qubits, and distance \(d\).

There are several different noise models used to evaluate the performance of quantum error correcting codes and decoders. 
The simplest model (often called \textit{code capacity}) assumes that the measurement of stabilizer operators can be done noiselessly, and considers a single application of some depolarizing channel to each of the data qubits in the code. 
This model is commonly used to evaluate the performance of machine learning decoders (\textit{e.g.}~\cite{torlai2017neural,krastanov2017deep,andreasson2019error,liu2019neural,ni2020neural,varsamopoulos2020decoding,colomer2020reinforcement,wagner2020symmetries,cao2023qecgpt,gong2024graph,gicev2023scalable,maan2024machine,ninkovic2024decoding}). 

However, in an actual quantum computer, stabilizer operators must be measured using some circuit, the components of which will all be faulty. 
This can result in the observed measurement results being incorrect, as well as single circuit errors propagating to multiple qubits in the code. 
An error model that applies a independent depolarizing channel to every possible faulty location in the circuit is called a \textit{circuit-level} noise model and is a more accurate method of evaluating a code and decoder (although it still falls short of realistically representing all of the sources of noise present in an actual quantum computer).

Oftentimes, both code capacity and circuit-level noise models are parametrized by a single value \(p\), from which the error rate of each specific kind of noise is derived.
For example, in a simple code capacity noise model, each qubit experiences an \(X\), \(Y\), or \(Z\) error, each with probability \(p/3\). Given a quantum error correcting code, such a noise model, and a decoder, it is possible to determine the \textit{psuedo-threshold} -- a value \(p^*\), such that for \(p < p^*\), we have \(P_L < p\), where \(P_L\) is the logical error rate.
The psuedo-threshold can be thought of as the break-even point, below which using the quantum error correcting code is helpful in combating noise.

\subsection{Transformers}
\label{sec:transformers}

We now introduce the building blocks of our transformer-based ML model - the attention mechanism and positional encodings.
This section focuses on a simple description of the attention mechanism and some common variants, as well as what positional encodings are and why they are needed. 
How these elements are used in our model is described in Section~\ref{sec:model-architecture}, and the details of code-aware self-attention, a variant of attention designed specifically for the task of error correction, are given in Section~\ref{sec:code-aware-self}.

Transformers, originally introduced in~\cite{vaswani2017attention}, are a class of deep neural networks that have found success in a wide variety of tasks, most notably in natural language processing with large language models~\cite{naveed2024comprehensive}.
A key aspect of the transformer architecture is the attention mechanism, which can be thought of as soft lookup-table, in which the elements of the table have an associated \textit{key} and \textit{value}.
Given a \textit{query}, the match with every key in the table is computed, and then used to update the query as a linear combination of the values.

More concretely, consider a sequence of \(d_m\)-dimensional vectors \((\mathbf{x}_0, \dots, \mathbf{x}_N)\).
A self-attention mechanism learns three linear maps from \(\mathbb{R}^{d_m}\) to \(\mathbb{R}^{d_k}\), which are labeled \(Q, K\), and \(V\) (standing for queries, keys, and values).
We then compute the attention matrix \(\alpha\), with entries
\begin{equation}
\label{eq:attn1}
\alpha_{i,j} = \frac{e^{\beta_{i,j}}}{Z_i}
\end{equation}
where
\begin{equation}
\label{eq:attn2}
\beta_{i,j} = \frac{\langle Q(\mathbf{x}_i), K(\mathbf{x}_j)\rangle}{\sqrt{d_k}}
\end{equation}
and 
\begin{equation}
\label{eq:attn3}
Z_i = \sum_{j=1}^{N} e^{\beta_{i,j}}
\end{equation}
where in Equation \eqref{eq:attn2} we use \(\langle \cdot, \cdot \rangle\) to denote the standard inner product
(note that the \(\alpha\) matrix is obtained from the \(\beta\) matrix by taking the \textrm{softmax} over the rows of the \(\beta\) matrix). Then, the updated vectors in the sequence are computed as
\begin{equation}
  \label{eq:attn4}
\mathbf{x}_i' = W \sum_j \alpha_{i,j} V(\mathbf{x}_j).
\end{equation}
where \(W\) is a learned \(d_m \times d_k\) matrix. Loosely, we can think of the elements of \(\alpha\) matrix as how much ``attention'' the model should pay to vector \(\mathbf{x}_j\) when updating \(\mathbf{x}_i\).

Transformer models (including the ones used in this work) often use multi-headed attention. In multi-headed attention,
there are multiple versions of each of the \(Q, K\), and \(V\) maps, which we denote by \(Q^{(h)}, K^{(h)}\), and \(V^{(h)}\), for
\(h \in \left\{ 1, \dots, H \right\}\).
\(H\) is referred to as the number of heads.
Each attention head computes an independent set of attention weights, \(\alpha_{i,j}^{(h)}\), and then equation~\eqref{eq:attn4} is changed to be
\begin{equation}
\label{eq:attn5}
\mathbf{x}_i' = \sum_{h=1}^{H}W^{(h)} \sum_j \alpha_{i,j}^{(h)} V^{(h)}(\mathbf{x}_j).
\end{equation}

An additional form of the attention mechanism is \textit{cross-attention}, where there are two sequences \(\{\mathbf{x}_i\}\) and \(\{\mathbf{y}_j\}\).
One sequence is used to compute the queries, (\(Q(\mathbf{x}_i)\)), while the others are used to compute the keys (\(K(\mathbf{y}_j)\)) and values (\(V(\mathbf{y}_j)\)).

In the prior description of attention, we have referred to the inputs and outputs as a ``sequence'' of vectors, even though equations~\eqref{eq:attn1}-\eqref{eq:attn5} make no use of any sequential structure. 
However, in many situations the order of the input vectors is important. 
When this is the case, this information can be added to the data via a positional encoding - a map \(\mathrm{pos}: \left\{ 1, \dots, N \right\} \to \mathbb{R}^{d_m}\) (for an input sequence of length \(N\)). 
The inputs are then updated as \(\mathbf{x}_i' = \mathbf{x}_i + \mathrm{pos}(i)\).
Positional encodings can either be hand-crafted (as was done in~\cite{vaswani2017attention}) or learned (as is done in this work).

\subsection{Decoding Task}
\label{sec:decoding-task}

In this section, we consider the task of decoding circuit-level noise for circuits made of Clifford gates and Pauli measurements. 
We begin by formulating the task in such a way that an ML model that learns to perform the task well will approximate a maximum likelihood decoder.
We then describe how the task differs from the problem solved by other decoders such as BP-OSD, and show how a decoder that solves this task is still sufficient for fault-tolerant quantum computing.

Suppose we have a physical Clifford circuit \(C\) which consists of some number of rounds, \(N_R\).
For each round \(t\), the circuit measures a set of stabilizer operators for some stabilizer code \(\mathcal{S}_t\).
We do not require that \(\mathcal{S}_t = \mathcal{S}_{t'}\) for \(t \ne t'\), and we allow for arbitrary Clifford gates and measurement of Pauli operators between rounds of measuring stabilizer operators.

A \textit{detector} is some combination of check measurements that has a deterministic parity in the absence of noise.
For example, in a memory experiment which consists of measuring the same set of stabilizer operators repeatedly, two consecutive measurements of the same check operator constitute a detector.
For our purposes, a \emph{logical measurement} is some set of physical measurements whose product corresponds to the value of a logical operator. We assume that all of the physical measurements going into a logical measurement occur in-between rounds of check measurements.
Note that logical measurements may or may not have a known parity in the absence of noise.
Suppose there are \(N_D\) detectors in the circuit, and \(N_L\) logical measurements.

We consider some independent error model
\begin{equation}
 \mathcal{E} = \left\{ \mathcal{E}_{j, m} \right \}, \quad \mathcal{E}_{j, m} = \left( e_{j, m}, p_{j,m} \right)
\end{equation}
where \(j \in \left \{0, 1, \dots |C|-1 \right \}\) indexes over locations in the circuit, \(m\) indexes over elements of the \(n_j\) qubit Pauli group (where the number of qubits depends on the circuit location), \(e_{j,m} \in \mathcal{P}_{n_j}\), and \(p_{j,m} \in \left[ 0, 1 \right]\) is the probability of the error. 

Note that in addition to indexing errors by a circuit location \(j\) and Pauli group element \(m\), we can also use a single index \(l \in \{0, 1, \dots, |\mathcal{E}| - 1\}\).
We will switch back and forth between these two indexing conventions, using two indices when we wish to emphasize that \(e_{j, m}\) is an element of the \(n_j\) qubit Pauli group, and \(\mathcal{E}_l\) when we wish to emphasize that it is an element of a set of size \(|\mathcal{E}|\).

Since the global phase of the errors is unobservable, we only care about errors \(e_{j, m} \in P_{n_j} / \left\{ \pm 1, \pm i \right\}\).
Thus, for each error, we have that \(e_{j, m}^2 = I\). 
This allows us to associate each error \(\mathcal{E}_l\) with a vector \(\mathbf{e}_{l} \in \mathbb{F}_2^{N_E}\), where \({(\mathbf{e}_l)}_k = \delta_{lk}\) and \(N_E = |\mathcal{E}|\).
This mapping does not capture all of the structure of this set, \emph{e.g.} if there is a single qubit location where \(e_{j, m_1} = X, e_{j,m_2} = Y\), and \(e_{j, m_3} = Z\), we have that \(e_{j, m_1} e_{j, m_2} = e_{j, m_3}\) (up to phase), but \(\mathbf{e}_{j, m_1} + \mathbf{e}_{j, m_2} \ne \mathbf{e}_{j, m_3}\).

We define two matrices, \(D \in \mathbb{F}_2^{N_D \times N_E}\), and \(L \in \mathbb{F}_2^{N_L \times N_E}\), where

\begin{align}
 D_{jk} &= \begin{cases} 1 &\text{ if error } k \text{ flips detector } j \\ 0 &\text{ otherwise} \end{cases} \label{eq:dmatrix} \\
L_{jk} &= \begin{cases} 1 &\text{ if error } k \text{ flips logical measurement } j \\ 0 &\text{ otherwise} \end{cases} \label{eq:lmatrix}.
\end{align}
The matrix \(D\) can be thought of as the biadjacency matrix of a ``space-time Tanner graph'', where there is an edge between error node \(k\) and detector node \(j\) if and only if \(D_{jk} = 1\).
Furthermore, given the \(D\) and \(L\) matrices we can adjust the error model \(\mathcal{E}\), combining two error mechanisms when they flip the same set of detectors and logical measurements.
The probability of the combined error is given by the probability that only one of the two original errors occurred, \(p = p_1(1-p_2) + p_2(1-p_1)\).

When we perform the circuit, we will get some set of outcomes for the detectors that we represent with a vector \(\mathbf{d} \in \mathbb{F}_2^{N_D}\), and we will get some set of outcomes for the logical measurements that we represent as \(\mathbf{l} \in \mathbb{F}_2^{N_L}\).
If some set of errors \(\{\mathcal{E}_j\}, j \in J\) (\(J\) is just some indexing set) occur during the circuit, then \(\mathbf{d} = D \mathbf{e}\) where \(\mathbf{e} = \sum_{j \in J} \mathbf{e}_j\).

When decoding, we want to determine \(L \mathbf{e}\), which we refer to as the \textit{logical measurement flips}, so we can adjust our recorded logical measurements.
The task of the decoder is then to find \(\hat{\mathbf{e}}_{L}\), which we refer to as the predicted logical measurement flips, where
\begin{equation}
\label{eq:dec-task}
\hat{\mathbf{e}}_L = \argmax_{\mathbf{e}_L \in \mathbb{F}_2^{N_L}} p \left( \mathbf{e}_L | \mathbf{d} \right)
\end{equation}
and
\begin{equation}
\label{eq:prob}
p(\mathbf{e}_L | \mathbf{d}) = \sum_{\substack{\mathbf{e} \in \mathbb{F}_2^{N_E} \text{s.t.} \\ D\mathbf{e} = \mathbf{d} \\ L\mathbf{e} = \mathbf{e}_L}} p(\mathbf{e}).
\end{equation}
In words, the task is to determine the most likely corrections that need to be made to the logical measurement outcomes, given the observed detection events. 
We note that equation~\eqref{eq:prob} is equivalent to summing over the cosets of \(\mathbb{F}_2^{N_E} / A_E\), where \(A_E = \text{ker}(D) \cap \text{ker}(L)\).

Note that the goal of this task -- to produce the most likely logical measurement flips -- is different than the goal of a decoder such as BP-OSD, which is to provide a likely circuit-level error that can either be tracked as a Pauli-frame or used to apply a correction.
However, predicted logical measurement flips are sufficient for correcting the output of a circuit consisting of Clifford gates and Pauli measurements.

For many quantum error correcting codes, the ability to perform physical Clifford gates and Pauli measurements is sufficient for performing logical Clifford gates and Pauli measurements.
The model of quantum computation in which circuits are limited to Clifford gates and Pauli measurements is known as \textit{Pauli-Based Computation} (PBC)~\cite{bravyi2016trading}, and is universal for quantum computation when supplied with sufficiently high-fidelity magic states.
Additionally, PBC is a promising approach towards practical fault-tolerant computation. 
Several prominent surface code architectures~\cite{fowler2019low,litinski2019game,Chamberland2022universal,litinski2022active} envision large-scale quantum computation with PBC and magic state distillation~\cite{bravyi2005universal}.
For QLDPC codes, recent works~\cite{xu2024fast,cross2024improved} proposed specialized schemes of logical Pauli measurements to implement PBC on specific codes.\footnote{In particular, on the $[[144,12,12]]$ BB code, the work of~\cite{cross2024improved} showed how to perform the full logical Clifford group using logical measurements and automorphism gates.}
The work of~\cite{he2025extractors} further proposed the extractor architectures, which enable parallelized PBC on arbitrary QLDPC codes. Such flexibility makes the extractor architectures highly optimizable and scalable as a practical proposal for QLDPC-based quantum computers.
With these in mind, decoders that can rapidly and accurately predict logical measurement faults induced by physical errors in these circuits have valuable applications in practical fault-tolerance.

Finally, we note that similar formulations of decoding have been proposed before,~\cite{bombin2023modular,barber2025real,higgott2025sparse}, and we would like to continue to emphasize that it is sufficient for logical computation in the context of PBC, provided that the decoder is capable of producing the elements of the \(\hat{\mathbf{e}}_L\) vector in real time as they are needed.

\subsection{Model Architecture}
\label{sec:model-architecture}

We attempt to solve the problem given in equation~\eqref{eq:dec-task} by learning the conditional density \(p(\mathbf{e} \mid \mathbf{d})\) using a transformer based neural network (we now drop the subscript \(L\) in \(\mathbf{e}_L\) for clarity).
We first show how to express the problem of learning the conditional density as a maximum likelihood estimation (MLE) task, and then describe the specific model we use in order to perform MLE. We note that we are now restricting our attention to memory circuits, that is, we assume that the same set of check operators is measured some number of times.

First, we follow~\cite{cao2023qecgpt} and~\cite{fakoor2020trade} and factorize the density as
\begin{equation}
\label{eq:1}
p(\mathbf{e} \mid \mathbf{d} ) = \prod_{i=1}^{N_{L}} p_i(e_i \mid \mathbf{e}_{<i}, \mathbf{d}).
\end{equation}
where \(\mathbf{e}_{<i} = (e_1, e_2, \dots, e_{i-1})\), and \(i\) indexes the logical measurement flips. 

Our goal is to train a model \(f\), with parameters \(\theta\), such that 
\begin{equation}
f \left (\mathbf{e}_{<i}, \mathbf{d}; \theta \right )
\approx p_i\left (e_i = 1 \mid \mathbf{e}_{<i}, \mathbf{d} \right ).
\end{equation}

Following standard machine learning practice~\cite{prince2023understanding}, we attempt to learn the conditional distributions \(p_i\) by first picking a parameterized distribution \(q_i(e_i | \lambda_i)\).
We then pick a model with parameters \(\theta\) to output the values of \(\lambda_i\) given the inputs (in this specific case, \(\mathbf{e}_{<i}\) and \(\mathbf{d}\)). 
Finally, we perform maximum likelihood estimation to determine \(\theta\).

More concretely, the MLE is performed as follows.  We parameterize each conditional distribution as a Bernoulli distribution with success probability \(\lambda_i\): \(q_i(e_i | \lambda_i) = \lambda_i^{e_i}(1-\lambda_{i})^{1-e_i}\).
The model then predicts \(\lambda_i\).
During training, we maximize the likelihood of the
observed data \(\left\{ \mathbf{e}^{(j)}, \mathbf{d}^{(j)} \right\}\), that is, we wish to find model parameters \(\hat{\theta}\) where

\begin{align}
  \hat{\theta} &= \argmax_{\theta} \prod_j \prod_{i=1}^{N_L} q_i \left (e_i^{(j)} | f\left(\mathbf{e}_{<i}^{(j)}, \mathbf{d}^{(j)}; \theta \right)\right) \\
  &= \argmax_{\theta} \log \left( \prod_j  \prod_{i=1}^{N_L} q_i\left (
              e_i^{(j)} \mid f(\mathbf{e}_{<i}^{(j)}, \mathbf{d}^{(j)}; \theta )
            \right ) \right) \\
          &= \argmax_{\theta} \sum_j\sum_{i=1}^{N_L} \Big [e_i^{(j)} \log \left(
              f(\mathbf{e}_{<i}^{(j)}, \mathbf{d}^{(j)}; \theta )
            \right)\nonumber\\
            &+ (1 - e_i^{(j)} ) \log \left(
              1 - f(\mathbf{e}_{<i}^{(j)}, \mathbf{d}^{(j)}; \theta )
            \right)\Big] \label{eq:6}
\end{align}
where we have used the fact that taking the logarithm does not affect the \(\argmax\).
We can see that equation~\eqref{eq:6} is equivalent to minimizing the binary cross-entropy.
Assuming the model has indeed minimized the binary cross-entropy, we see that we can find the \(\argmax\) of equation~\eqref{eq:1} by
taking the predicted logical error vector \(\hat{\mathbf{e}}\) as
\begin{equation}
\label{eq:3}
\hat{e}_i = 
\begin{cases}
  1 & \text{ if } f \left (\mathbf{e}_{<i}, \mathbf{d}; \theta \right ) \ge \frac{1}{2} \\
  0 & \text{otherwise}
\end{cases}.
\end{equation}

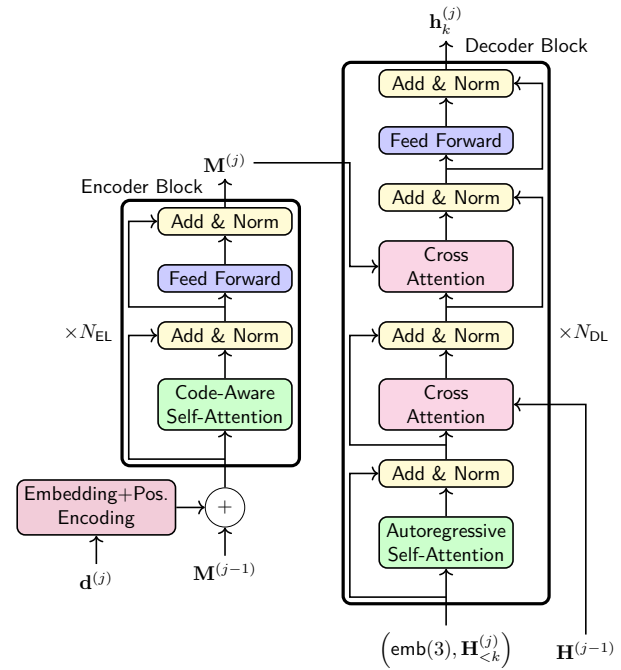
\begin{figure}[ht]
  \centering
  \resizebox{\columnwidth}{!}{
  {\fontfamily{lmss}\selectfont
    \begin{tikzpicture}[
    module/.style={draw, thick, rounded corners, minimum width=15ex},
    embmodule/.style={module, fill=purple!20},
    samodule/.style={module, fill=green!20},
    camodule/.style={module, fill=magenta!20},
    normmodule/.style={module, fill=yellow!20},
    ffmodule/.style={module, fill=blue!20},
    ]
    \node (inputs) {\(\mathbf{M}^{(j-1)}\)};
    \node[above=0.5 of inputs, draw, circle] (add) {\(+\)};
    \node[left=0.5 of add, embmodule, align=center] (inputemb) {Embedding+Pos.\\Encoding};
    \node[below=0.5 of inputemb] (inputs2) {\(\mathbf{d}^{(j)}\)};
    \node[above=1.0 of add, samodule, align=center] (selfattn) {Code-Aware\\Self-Attention};
    \node[above=0.5 of selfattn, normmodule] (norm1) {Add \& Norm};
    \node[above=0.5 of norm1, ffmodule] (ffnn) {Feed Forward};
    \node[above=0.5 of ffnn, normmodule] (norm2) {Add \& Norm};
    \node[above=0.5 of norm2] (output) {\(\mathbf{M}^{(j)}\)};

    \draw[->, thick] (inputs) -- (add);
    \draw[->, thick] (inputs2) -- (inputemb);
    \draw[->, thick] (inputemb) -- (add);
    \draw[->, thick] (add) -- (selfattn);
    \draw[->, thick] (selfattn) -- (norm1);
    \draw[->, thick] (norm1) -- (ffnn);
    \draw[->, thick] (ffnn) -- (norm2);
    \draw[->, thick] (norm2) -- (output);

    \coordinate (msaresidual) at ($(add.north)!0.5!(selfattn.south)$);
    \coordinate (ffnnresidual) at ($(ffnn.south)!0.5!(norm1.north)$);
    \coordinate[left=0.5 of norm1] (lnorm1);
    \coordinate[left=0.5 of norm2] (lnorm2);
    \coordinate[left=0.8 of output] (topleft);
    \coordinate (labelloc) at ($(lnorm2.north)!0.6!(topleft.south)$);
    \draw[->, thick] (msaresidual) -| (lnorm1) -- (norm1);
    \draw[->, thick] (ffnnresidual) -| (lnorm2) -- (norm2);

    \node[fit={(selfattn)(norm2)(lnorm2)(msaresidual)}, draw, ultra thick, rounded corners, label=left:\(\times N_{\text{EL}}\)] (encoder) {};
    \node (label) at (labelloc){Encoder Block};

    \node[right=1.5 of selfattn, camodule, align=center] (crossattn1) {Cross\\Attention};
    \node[below=0.5 of crossattn1, normmodule] (norm3) {Add \& Norm};
    \node[below=0.5 of norm3, samodule, align=center] (aselfattn) {Autoregressive\\Self-Attention};
   
     \node[below=1.0 of aselfattn] (inputs2) {\(\left(\text{emb}(3), \mathbf{H}_{<k}^{(j)}\right)\)};
    \node[above=0.5 of crossattn1, normmodule] (norm4) {Add \& Norm};
    \node[above=0.5 of norm4, camodule, align=center] (crossattn2) {Cross\\Attention};
    \node[above=0.5 of crossattn2, normmodule] (norm5) {Add \& Norm};
    \node[above=0.5 of norm5, ffmodule] (ffnn2) {Feed Forward};
    \node[above=0.5 of ffnn2, normmodule] (norm6) {Add \& Norm};
    \node[above=0.5 of norm6] (output2) {\(\mathbf{h}_k^{(j)}\)};

    \node[right=0.5 of inputs2] (inputs3) {\(\mathbf{H}^{(j-1)}\)};

    \draw[->, thick] (inputs2) -- (aselfattn);
    \draw[->, thick] (aselfattn) -- (norm3);
    \draw[->, thick] (norm3) -- (crossattn1);
    \draw[->, thick] (crossattn1) -- (norm4);
    \draw[->, thick] (norm4) -- (crossattn2);
    \draw[->, thick] (crossattn2) -- (norm5);
    \draw[->, thick] (norm5) -- (ffnn2);
    \draw[->, thick] (ffnn2) -- (norm6);
    \draw[->, thick] (norm6) -- (output2);

    \coordinate[left=0.5 of crossattn2] (lca2);
    \coordinate[left=0.5 of norm3] (lnorm3);
    \coordinate[left=0.5 of norm4] (lnorm4);
    \coordinate[right=0.5 of norm5] (rnorm5);
    \coordinate[right=0.5 of norm6] (rnorm6);
    \coordinate[right=0.5 of crossattn1] (rcrattn1);
    \coordinate (asaresidual) at ($(inputs2.north)!0.5!(aselfattn.south)$);
    \coordinate (craresidual1) at ($(crossattn1.south)!0.5!(norm3.north)$);
    \coordinate (craresidual2) at ($(crossattn2.south)!0.5!(norm4.north)$);
    \coordinate (ffnnresidual2) at ($(ffnn2.south)!0.5!(norm5.north)$);
    \coordinate[right=0.8 of output2] (routput2);
    \coordinate (decoderlabel) at ($(routput2.south)!0.4!(rnorm6.north)$);

    \draw[->, thick] (asaresidual) -| (lnorm3) -- (norm3);
    \draw[->, thick] (craresidual1) -| (lnorm4) -- (norm4);
    \draw[->, thick] (craresidual2) -| (rnorm5) -- (norm5);
    \draw[->, thick] (ffnnresidual2) -| (rnorm6) -- (norm6);
    \draw[->, thick] (output) -| (lca2) -- (crossattn2);
    \draw[->, thick] (inputs3) |- (rcrattn1) -- (crossattn1);

    \node[fit={(norm6)(asaresidual)(lnorm3)(rnorm5)}, draw, ultra thick, rounded corners, label=right:\(\times N_{\text{DL}}\)] (decoder) {};
    \node at (decoderlabel){Decoder Block};

  \end{tikzpicture}
  }}
\caption{
  The encoder and decoder blocks of the machine learning decoder.
  In our model, for each round of syndrome measurement, the encoder block takes in the sum of (a) the output of the previous encoder block and (b) a learned embedding of the detector outcomes.
  Information about the code and syndrome measurement circuit is encoded in the \textit{code-aware self-attention} layer (see Section~\ref{sec:code-aware-self}).
  The decoder block takes in the output of the encoder block, the output of the previous decoder block, and then produces some number of \textit{latent space
  predictions} \(\mathbf{h}\).
  Not shown in the figure, the last application of the decoder block is followed by a linear layer and a sigmoid function, the output of which is then used to make
  predictions of the logical measurement flips.}
  \label{fig:model-arch2}
\end{figure}

A diagram of the model we use is shown in Figure~\ref{fig:model-arch2}.
Here, we provide an overview of the model and defer some of the details to appendix~\ref{sec:model-arch-deta}.
The model consists of two blocks, an encoder block \(\mathsf{E}\) and a decoder block \(\mathsf{D}\) (where we use ``encoder'' and ``decoder'' in the sense of the standard transformer architecture, not in the sense of QEC).
For each round of syndrome measurement in the circuit, we will perform one iteration both the encoder and decoder blocks.

The encoder block takes in one input, which is the sum of a learned embedding
\footnote{An \emph{embedding} refers either to a mapping from some high dimensional input space, such as the space of possible detector outcomes, to some usually lower dimensional output space such as \(\mathbb{R}^n\), or a specific output of such a mapping.}
of the detector outcomes from a round of syndrome measurement and the output of the previous iteration of the encoder block.
We pick the output space of the embedding function to be the same size as the input and output spaces of the encoder block.
We represent the action of the encoder block for syndrome measurement round \(j\) as
\begin{equation}
\mathbf{M}^{(j)} = \mathsf{E} \left(\mathbf{M}^{(j-1)} + \text{emb}\left(\mathbf{d^{(j)}}\right) \right)
\end{equation}
where \(\text{emb}(\mathbf{d}^{(j)})\) is a learned embedding of the detector outcomes from round \(j\), and \(\mathbf{M}^{(j-1)}\) is the output of the previous application of the encoder block.
We include a positional encoding in the \(\text{emb}\) function.
For the first round, we take \(\mathbf{M}^{{0}}\) to be the learned embedding of all trivial detector outcomes.
Intuitively, our goal will be for the encoder to produce some learned representation \(\mathbf{M}^{(j)}\) that will encode the information of the detector outcomes from round \(j\), as well as detector outcomes from previous rounds. The encoder block is implemented as multiple rounds of masked self-attention and feedforward layers, as depicted in Figure~\ref{fig:model-arch2}.

The decoder block takes in three inputs.
These inputs are a special learned vector used to indicate what kind of output the decoder block is supposed to produce, the output from the encoder, and the output from the previous iteration of the decoder block.
For all but the last round of syndrome measurement, we then apply the decoder block \(c\) times (for some chosen value of \(c\)), producing \(c\) vectors \(\mathbf{h}_k^{(j)}\) (which we collectively refer to as a matrix \(\mathbf{H}^{(j)}\)).
We can represent the action of the decoder producing the \(k\)th vector as

\begin{equation}
\mathbf{h}_k^{(j)} = \mathsf{D}\left ( \left(\text{emb}(3), \mathbf{H}_{<k}^{(j)} \right), \mathbf{M}^{(j)}, \mathbf{H}^{(j-1)}  \right)
\end{equation}
where \(\text{emb}(3)\) is a learned embedding of the value \(3\), used to indicate to the decoder block to start generating \(\mathbf{h}\) vectors (as opposed to predictions of logical measurement flips, as it will do after the last round of syndrome measurement).
The number \(3\) can be thought of as analogous to the \textit{start of sentence} token often used in natural language processing models.
We refer to the \(\mathbf{h}\) vectors as \textit{latent space predictions}.
For the first round, we set \(\mathbf{H}^{(0)}\) to be an embedding of all trivial logical measurement flips.

For the last round of syndrome measurement, we instead apply the decoder block \(N_L\) times, as well as add a final feedforward layer, sigmoid function and step function to predict the logical measurement flips.
We can represent this as
\begin{align}
  \mathbf{h}_k &= \mathsf{D}\left(  \left( \text{emb}(2), \text{emb}\left(\hat{\mathbf{e}}_{<i}\right) \right),\mathbf{M}^{(N_R)}, \mathbf{H}^{(N_R-1)} \right) \label{eq:dec-block1}\\
\hat{e}_k &= u \left( \sigma \left( \mathbf{r}^T \mathbf{h}_k \right)  - 0.5 \right) \label{eq:dec-block2}
\end{align}
where \(u\) is the unit step function, \(\sigma\) is the sigmoid function, \(\text{emb}(2)\) is a learned embedding of the value \(2\), used to indicate to the decoder to generate logical measurement flip predictions, and \(\mathbf{r}\) is a learned vector used to project \(\mathbf{h}_k\) down to a scalar.

As shown in Figure~\ref{fig:model-arch2}, the decoder block is implemented as a layer of self-attention, and two layers of cross attention.
In the first cross-attention layer, the alternate sequence of vectors is the outputs of the prior decoder block, and in the second cross-attention layer, the alternate sequence of vectors is the output of the encoder block.

We note that our model can be thought of as performing sliding-window decoding (originally introduced as the "overlapping recovery method" in~\cite{dennis2002topological}) with a window size of one.
Indeed, as we describe in Section~\ref{sec:training}, we initially train the model exactly as a sliding-window decoder, which outputs predicted errors at the end of every syndrome measurement round.
For quantum error correcting codes that are not single-shot~\cite{bombin2015single}, such as BB codes, a window size of one is not sufficient to make accurate predictions, as the syndrome information is unreliable.
This is typically handled in sliding window decoding by simple using larger window sizes~\cite{skoric2023parallel,bombin2023modular,tan2023scalable,huang2023improved}, at the cost of increasing the runtime of the decoder.

Despite only using a window size of one, our model can still function as an accurate decoder by making latent space predictions, which allows the model to output information related to predicted errors, but does not force the model to commit to any given error before it has enough syndrome data.
This method of latent space predictions was inspired by a recently proposed paradigm for training large language models to perform reasoning tasks~\cite{hao2024training}.
We comment further on the latent space predictions and their affect on training in Section~\ref{sec:training}.

\subsection{Training}
\label{sec:training}
As described in Section~\ref{sec:model-architecture}, using latent space predictions allows the model to make accurate predictions while only consuming one round of syndrome measurement data at a time.
However, we find that when we attempt to directly train the model as described in Section~\ref{sec:model-architecture} on circuit-level noise for BB codes, the loss does not appreciable decrease.
To overcome this challenge, we adopt the multi-stage training protocol originally introduced in~\cite{hao2024training}, which was the original inspiration for the use of latent space predictions.

To describe this protocol in the context of decoding quantum error correcting codes, we first define \textit{intermediate logical measurement flips}. Let \(\mathbf{e}_C \in \mathbf{F}_2^{N_E}\) be the circuit-level error that occurred, and let \(\Pi_j\) be a projection onto the set of error locations from the first \(j\) rounds of syndrome measurement.
We then set \(\mathbf{e}^{(j)} = L \Pi_j \mathbf{e}_C\) to be the intermediate logical measurement flips after \(j\) rounds of syndrome measurement, where \(L\) is the matrix defined in equation~\eqref{eq:lmatrix}.

At a high level, the goal of this training procedure is to break the overall decoding task down into smaller steps that the model can more easily learn -- specifically, predicting intermediate logical measurement flips.
This is analogous to the \textit{chain of thought} prompting technique~\cite{wei2022chain} used to improve the ability of large language models to perform reasoning tasks.
By providing these easier, intermediate steps, we find that the training process becomes tractable.
However, because our model only processes a single round of syndrome information at a time, it is not actually able to produce intermediate logical measurement flip predictions with the accuracy we desire.
Thus, during training, we gradually transition away from making intermediate logical measurement flip predictions, to latent space predictions, until finally the only actual logical measurement flip predictions are made after the final round of syndrome measurement. 

We now provide the details on how this procedure is implemented. During the first stage of training, we use the decoder block to predict intermediate logical measurement flips after every round of syndrome measurement, which we represent by modifying equations~\eqref{eq:dec-block1} and~\eqref{eq:dec-block2} as
\begin{align}
  \mathbf{h}_k^{(j)} &= \mathsf{D}\left (
       \left(\text{emb}(2), \text{emb}\left(\hat{\mathbf{e}}^{(j)}_{<k}\right) \right),
      \mathbf{M}^{(j)},
      \text{emb}\left(\hat{\mathbf{e}}^{(j-1)}\right)
                       \right) \\
  \hat{e}_k^{(j)} &= \theta\left(\sigma \left(\mathbf{r}^T \mathbf{h}_k^{(j)}\right) - 0.5 \right).
\end{align}

Furthermore, during this stage of training, the loss function is modified from equation~\eqref{eq:6} to be the binary cross-entropy not just between the outputs of the last decoder block and the observed logical measurement flips, but between the outputs from every decoder block and the intermediate logical measurement flips as well.
Concretely, for a single training example \((\mathbf{e}, \mathbf{d})\), the loss is 
\begin{align}
\label{eq:loss1}
    \mathcal{L} = \sum_{j=1}^{N_R} &\sum_{k=1}^{N_L} e_k^{(j)} \log \left ( \sigma \left ( \mathbf{r}^T \mathbf{h}_k^{(j)} \right) \right ) \nonumber \\
    &+ \left (1 - e_k^{(j)}\right) \log \left ( 1 - \sigma \left ( \mathbf{r}^T \mathbf{h}_k^{(j)} \right) \right ).
\end{align}
We note that while in equation~\eqref{eq:6}, \(j\) indexes over examples in the training dataset, here, \(j\) indexes over rounds of syndrome measurement for a single training example.

During the next stage of training, we change the first decoder block to produce the more flexible latent space predictions instead of predictions of intermediate logical measurement flips.
The last \(N_R - 1\) rounds remain the same, and the binary cross-entropy loss is calculated using the outputs of the decoder block from the last \(N_R - 1\) rounds, that is, the outer sum runs from \(j=2\) to \(j=N_R\).

In subsequent stages of training, we continue to decrease the number of rounds for which the decoder outputs intermediate predicted logical measurement flips, modifying the loss function at each stage accordingly, until the decoder makes latent space predictions in all rounds except for the final round. 
By allowing the decoder to make predictions in latent space, our goal is to avoid forcing the model to commit to certain results without having enough information to do so.

\subsection{Code-Aware Self-Attention}
\label{sec:code-aware-self}

In this section we describe code-aware self-attention.
We first motivate the need for including code-specific information into the structure of the model, before describing how this is accomplished via code-aware self-attention, and how it is implemented in our model.
Finally, we compare the performance of training our model with and without code-aware self-attention.

Our model, as described in Section~\ref{sec:model-architecture}, makes no use of any structure particular to a specific error correcting code, syndrome measurement circuit, or noise model.
As noted in Section~\ref{sec:intro}, ML decoders that have performed well on larger distances surface codes have tended to take advantage of the local structure of the check operators.
In general, encoding assumptions about the structure of the data into the design of the model (a so-called \textit{inductive bias}) is thought to be an essential component of how ML models generalize to data beyond that in the training set~\cite{goyal2022inductive,prince2023understanding}.

While the check operators of a QLDPC code are no longer local in space, they are restricted to acting on a constant number of qubits.
Thus, using all-to-all self-attention in the encoder block will result in comparing detectors outcomes that are only weakly related to each other, as only high-weight errors touch both detectors.
This observation was made in~\cite{choukroun2022error} in the context of classical error correcting codes, and they developed code-aware self-attention as a way of restricting the attention mechanism to only consider attention between bits of the syndrome and bits of the codeword that are known to be highly correlated due to the structure of the parity check matrix.

\begin{figure}[htbp]
  \centering
  \includegraphics[width=0.9\columnwidth]{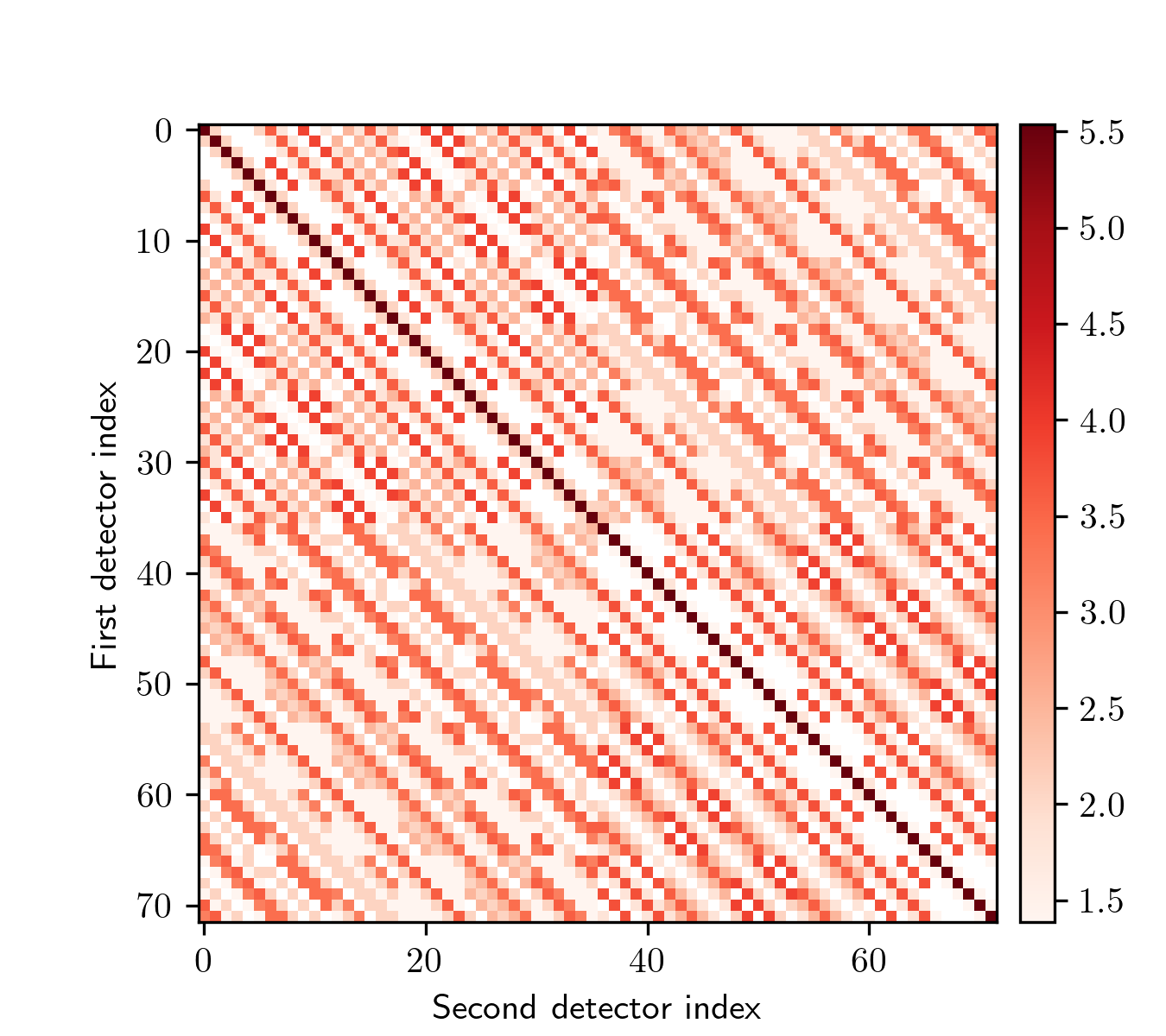}
  \caption{An example of a code-aware attention mask for the
    \([[72,12,6]]\) BB code. The value of the mask at position
    \(i, j\) is \(\log(n)\), where \(n\) is number of errors that flip both detectors \(i\) and \(j\). We note that this mask is symmetric, \textit{i.e.} the value at position \((i,j)\) is equal to the value at position \((j, i)\).}
  \label{fig:camask}
\end{figure}

To implement code aware-self attention, we adjust the standard attention equation (equations~\ref{eq:attn1}-\ref{eq:attn3}) as
\begin{equation}
\beta_{i,j} = \frac{\langle Q(\mathbf{x}_i), K(\mathbf{x}_j)\rangle}{\sqrt{d_k}} + M_{i,j}
\end{equation}
where the matrix \(M\) is called a \textit{mask}.
We modify the approach of~\cite{choukroun2022error} slightly, and take \(M = \log \left( D D^T \right)\) with matrix \(D\) as defined in equation (\ref{eq:dmatrix}), where we define \(\log(0) = - \infty\) and \(e^{-\infty} = 0\).
The matrix element \((DD^T)_{i,j}\) is the number of length-two paths on the space-time Tanner graph between detectors \(i\) and \(j\).
Thus, in the encoder block, we force attention weight \(\alpha_{i,j}\) to be zero if detectors \(i\) and \(j\) are a distance greater than two away on the Tanner graph, and provide a slight boost to the value of \(\alpha_{i,j}\) if detectors \(i\) and \(j\) are flipped by many common errors.
In each application of the encoder block, we use the restriction of \(M\) to the rows and columns corresponding to detectors from that round of syndrome measurement.
An example of such a mask is shown in Figure~\ref{fig:camask}.

\begin{figure}[htbp]
    \centering
    \includegraphics{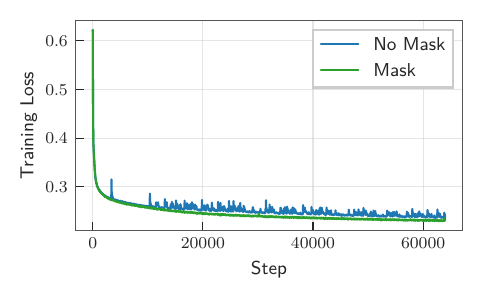}
    \caption{A comparison of the first stage of model training with and without code-aware self-attention.
    The blue loss curve ("No Mask") shows the results of training a model on a memory experiment with the \([[72,12,6]]\) BB code, using normal multi-headed self attention in the encoder block.
    The green loss curve ("Mask") shows the results of training the same model on the same memory experiment, but using code-aware self-attention as described in Section~\ref{sec:code-aware-self}.
    Both curves show a moving average of the training loss with a window size of \(50\) steps.}
    \label{fig:compare-mask}
\end{figure}

To evaluate the benefit provided by code-aware self-attention, we compare the loss curves of the model trained with and without the attention mask.
The loss curves are shown in Figure~\ref{fig:compare-mask}.
Each model was trained on a memory experiment with the \([[72,12,6]]\) BB code, with a physical error rate of \(p=0.6\%\).
The models were trained with the Adam optimizer~\cite{kingma2017adam} with a learning rate of \(10^{-4}\) using a batch size of \(512\).
For this experiment, we used zero rounds of latent space predictions.
The model hyperparameters can be found in Table~\ref{tab:1}. 
As shown in the figure, using code-aware self-attention results both in more stable training, and lower training losses.

\subsection{Memory Experiment}
\label{sec:memory-experiment}

In this section we describe the details of the memory experiments we performed to evaluate the ML decoder.
The memory experiments consisted of the following steps:

\begin{enumerate}
\item Noiselessly prepare all physical qubits in the \(\ket{+}\) state.
\item\label{item:1} Perform one round of noiseless syndrome measurement.
\item\label{item:2} Perform \(N_R\) rounds of noisy syndrome measurement.
\item\label{item:3} Perform one round of noiseless syndrome measurement.
\item\label{item:4} Measure the code's logical \(X\) operators noiselessly.
\end{enumerate}

To measure the syndromes, we use the circuits described in~\cite{bravyi2024highthreshold}.
These circuits use one ancilla qubit per \(X\) and \(Z\) check operator in the code.
Measuring an \(X\) check requires initializing the ancilla qubit in the \(|+\rangle\) state, followed by six CNOT gates between the ancilla and the data qubits in the support of the check, and finally measuring the ancilla in the \(X\) basis.
Similarly, measuring a \(Z\) check requires initializing the ancilla qubit in the \(|0\rangle\) state, followed by six CNOT gates, and finally measuring the ancilla in the \(Z\) basis.
For more details on these circuits, we refer the reader to the Section ``Syndrome circuit'' in~\cite{bravyi2024highthreshold}.

We use a standard circuit-level depolarizing noise model parameterized by a single value \(p\), in which:
\begin{enumerate}
\item All qubit initializations in the \(\ket{0}\) state are followed by an \(X\) error with probability \(p\), and all qubit initializations in \(\ket{+}\) basis are followed by a \(Z\) error with probability \(p\). 
\item All single qubit gates are followed by an \(X\), \(Y\), or \(Z\) error, each with probability \(p/3\).
\item All idle qubit locations are followed by an \(X\), \(Y\), or \(Z\) error, each with probability \(p/3\).
\item All two qubit gates are followed by a non-trivial two-qubit Pauli error (\textit{i.e.} \( \left (\mathcal{P}_2 / \langle i\rangle \right) \setminus \left \{ I^{\otimes 2}\right \}\)),  where each error occurs with probability \(p/15\).
\item All measurements results are flipped with probability \(p\).
\end{enumerate}

All circuit simulations were performed with \texttt{Stim}~\cite{gidney2021stim}.
The decoder takes in all of the detector outcomes from the circuit, and then makes predictions of the logical measurement flips (\textit{i.e.} whether a circuit-level error that anti-commutes with a logical measurement has occurred).
Since the code starts in a logical \(\ket{+}^{\otimes k}\) state, logical measurement flips are equivalent to the logical measurement outcomes.
If the prediction from the decoder of the logical measurement flips differs from the logical measurement result for any logical qubit, it is considered a logical \(Z\) error.
Note that for BP-OSD, we obtain the predicted logical measurement flips \(\hat{\mathbf{e}}_L\) through the relationship \(\hat{\mathbf{e}}_L = L \hat{\mathbf{e}}\), where \(\hat{\mathbf{e}}\) is the predicted circuit level error from BP-OSD and \(L\) is the matrix defined in equation~\eqref{eq:lmatrix}, whereas the ML decoder directly outputs \(\hat{\mathbf{e}}_L\).

\section{Contributions and Acknowledgments}
IC conceived of and supervised the project. JB and IC developed the NN architecture with input from LZ and ZH. JB and HA wrote the training code and ran the experiments. JB, ZH, and IC wrote the manuscript. All authors contributed to scientific discussions and analysis of results.

The authors thank Andrew Cross, Ted Yoder, and Patrick Rall for helpful discussions throughout this project.
Z.H.~is supported by the MIT Department of Mathematics and the NSF Graduate Research Fellowship Program under Grant No. 2141064.
This work is supported in part by the National Science Foundation under Cooperative Agreement PHY-2019786 (The NSF AI Institute for Artificial Intelligence and Fundamental Interactions, \url{https://iaifi.org}) and the CQE-LPS Doc Bedard Fellowship.
This work made use of resources provided by subMIT at MIT Physics, as well as the FASRC Cannon cluster supported by the FAS Division of Science Research Computing Group at Harvard University.

\bibliographystyle{quantum}
\bibliography{bibliography}

\onecolumn
\appendix

\section{Model Architecture Details}
\label{sec:model-arch-deta}

In this section we give a detailed description of the model architecture, defining all of the input and output spaces for each piece of the model as well as all of the hyperparameters.
Our model (see Figure~\ref{fig:model-arch2}) consists of repeated applications of an encoder block and a decoder block, where the number of applications is equal to the number of rounds of syndrome measurement (\(N_R\)).
We make the following definitions:
\begin{itemize}
\item Let \(d_m \in \mathbb{N}\) be a natural number we refer to as the ``model
  dimension''.
\item Let \(\mathbf{d}^j \in \mathbb{F}_2^{N_D}\) be the detector outcomes from round \(j\) of syndrome measurement, for \(j \in \left\{ 1, \dots, N_R \right\}\).
\item Let \(\Pi_j: \mathbb{F}_2^{N_E} \to \mathbb{F}_2^{N_E}\) be a projector onto error mechanisms from the first \(j\) rounds of syndrome measurement.
We then define \(\mathbf{e}^j = L \Pi_j \mathbf{e}\) to be the logical measurement flips at the end of syndrome measurement round \(j\), where \(L\) is the logical matrix as defined in Section~\ref{sec:decoding-task}, and \(\mathbf{e} \in \mathbb{F}_2^{N_E}\) is the error that occurred in the circuit.
\item Let \(\text{emb}_d: \left\{ 0, 1 \right\} \to \mathbb{R}^{d_m}\) be a function that represents the embedding of detector outcomes into \(\mathbb{R}^{d_m}\), and let \(\text{emb}_e: \{0, 1, 2, 3\} \to \mathbb{R}^{d_m}\) be an embedding of errors (along with two other ``tokens'' to be defined later) into \(\mathbb{R}^{d_m}\).
\item Let \(\text{pos}_d: \{1, \dots, N_D\} \to \mathbb{R}^{d_m}\) be the positional encoding for detection events, and \(\text{pos}_e: \{1, \dots, N_L\} \to \mathbb{R}^{d_m}\) be the positional encoding for logical measurement flips.
\item Let \(n_h \in \mathbb{N}\) be the number of heads.
\item Let \(N_{\text{EL}}, N_{\text{DL}} \in \mathbb{N}\) be the number of encoder and decoder layers, respectively.
\end{itemize}

The encoder block is a function \(\mathsf{E}: \mathbb{R}^{N_D \times d_m} \to \mathbb{R}^{N_D \times d_m}\), and it processes detector outcomes one round at a time, producing some representations
\(\mathbf{M}^{(j)} \in \mathbb{R}^{N_D \times d_m}\), \textit{i.e.} for
\(j \in \left\{ 1, 2, \dots, N_R \right\}\),
\begin{equation*}
  \mathbf{M}^{(j)} = \mathsf{E}\left(
    \mathbf{M}^{(j-1)} + \text{emb}_d\left(\mathbf{d}^{i-1}\right)
    + \text{pos}_d([1, \dots, N_D)
  \right)
\end{equation*}
where we set \(\mathbf{M}^{(0)} = \text{emb}_d([0, 0, \dots, 0]) + \text{pos}_d([1, 2, \dots, N_D])\) (the embedding of all trivial detection events).

As shown in Figure~\ref{fig:model-arch2}, the encoder block is implemented via \(N_{\text{EL}}\) layers, where each layer consists of multi-headed code-aware self-attention, followed by an application of layer normalization~\cite{ba2016layer}, followed by a feedforward layer, and finally another layer normalization.
We use \(n_h\) heads in the multi-headed code-aware self-attention layers.
We also include residual connections that go around the self-attention and feedforward layers.

The goal of the model is to predict the logical measurement flips at the end of round \(N_R\). 
To do this, we use the decoder block, which is a function \(\mathsf{D}: \mathbb{R}^{N \times d_m} \times \mathbb{R}^{N_D \times d_m} \times \mathbb{R}^{N' \times d_m} \to \mathbb{R}^{d_m}\) (that is, \(\mathsf{D}\) takes in three sequences of \(d_m\) dimensional vectors, the first sequence being length \(N\), the second being length \(N_D\), and the third being length \(N'\), and outputs a single \(d_m\)-dimensional vector) where \(N\) and \(N'\) are between \(1\) and \(N_L\).
For the first \(N_R - 1\) rounds, the decoder block produces \(c\) dimension \(d_m\) vectors, concretely,
for \(1 \le j \le N_R - 1\) and \(1 \le k \le c\),
\begin{equation*}
  \mathbf{h}^{(j)}_k = \mathsf{D}\left(
  [\text{emb}_e(3), \mathbf{H}^{(j)}_{<k}],
  \mathbf{M}^{(j)},
  \mathbf{H}^{(j-1)}
  \right).
\end{equation*}
where \(\mathbf{H}^{(j)} = [\mathbf{h}^{(j)}_1, \dots, \mathbf{h}^{(j)}_c]\), and \(\mathbf{H}^{(j)}_{<k} = [\mathbf{h}^{(j)}_1, \dots, \mathbf{h}^{(j)}_{k-1}]\).
For the first round, we set the last argument of \(\mathsf{D}\) to be
\begin{equation*}
\text{emb}_e([0, 0, \dots, 0]) + \text{pos}_e([0, 1, \dots, N_L-1]).
\end{equation*}

For the last round (\(j = N_R\)), the decoder is used to predict logical measurement flips \(\hat{\mathbf{e}}^{N_R}\), where for \(1 \le k \le N_L\),

\begin{equation*}
  \mathbf{h}_k^{(N_R)} = \mathsf{D}\left(
    \text{emb}_e ([2, \hat{\mathbf{e}}^{(N_R)}_{<k}])
    + \text{pos}_e([1, \dots, k-1]),
  \mathbf{M}^{(N_R)},
  \mathbf{H}^{(N_R - 1)}
  \right)
\end{equation*}
and
\begin{equation*}
  \hat{e}^{(N_R)}_k =
  \begin{cases}
    1 & \text{ if } \sigma \left(\mathbf{r}^T \mathbf{h}_k^{(N_R)}\right) \ge 0.5\\
    0 & \text{ otherwise }
  \end{cases}.
\end{equation*}
where \(\sigma\) is the sigmoid function and \(\mathbf{r} \in \mathbb{R}^{d_m}\).

As shown in Figure~\ref{fig:model-arch2}, the decoder block is implemented via \(N_{\text{DL}}\) layers, where each layer consists of multi-headed self-attention, two layers of multi-headed cross-attention, and a final feedforward network, with layer normalizations interspersed.
We note that in the self-attention layer, we used a mask to enforce the autoregressive property of the network, \textit{i.e.} we set \(\alpha_{ij} = 0\) for \(j > i\), such that the update to a query \(i\) can only depend on things earlier in the sequence.
We also used residual connections around the attention and feedforward layers.

Finally, we note that we included dropout layers with \(p=0.1\) after every attention layer in the network, and within each feedforward layer within the network.
We used the gelu activation function~\cite{hendrycks2023gaussian} in the feedforward layers for both the encoder and decoder.

\section{Training Details}
\label{sec:training-details}

In this section we provide the hyperparameters and additional training details of the models shown in Figures~\ref{fig:72-lers-times}, and~\ref{fig:72-12-6-multi}, \ref{fig:144-results}.
All models were trained with the Adam optimizer~\cite{kingma2017adam}.
Every epoch consisted of \(16,384\) samples generated using \texttt{Stim}.
We re-sampled a new set of \(16,384\) shots every epoch, essentially allowing us to work in the unlimited data regime.
In Table~\ref{tab:1} we give the hyperparameters of the models used to obtain the results on the \([[72,12,6]]\) code shown in Figures~\ref{fig:72-lers-times} and~\ref{fig:72-12-6-multi}.
Table~\ref{tab:3} shows the training hyperparameters used for these models.
In Table~\ref{tab:2} we give the hyperparameters of the model used to obtain the results on the \([[144,12,12]]\) code shown in Figure~\ref{fig:144-results}, and in Table~\ref{tab:4} we give the training hyperparemeters used for this model.

\begin{table}[ht]
\begin{center}
\begin{tabular}{ |c|c|c| }
  \hline
  Number of Encoder Layers & \(3\) \\ 
  Number of Decoder Layers & \(3\) \\  
  Number of Attention Heads (\(n_h\)) & \(8\) \\
  Model Dimension (\(d_m\)) & 256\\
  Feedforward Dimension (\(d_f\)) & 512 \\
  Number of latent space predictions (\(c\)) & 1\\
  Total number of parameters & \(4.77 \times 10^{6} \)\\
  \hline
\end{tabular}
\end{center}
  \caption{Hyperparameters of the model used to obtain the results shown in Figures~\ref{fig:72-lers-times} and~\ref{fig:72-12-6-multi}.
  Model weights are stored in single precision.}
\label{tab:1}
\end{table}

\begin{table}[ht]
\begin{center}
\begin{tabular}{ |c|c|c| }
  \hline
  Number of Encoder Layers & \(3\) \\ 
  Number of Decoder Layers & \(3\) \\  
  Number of Attention Heads (\(n_h\)) & \(8\) \\
  Model Dimension (\(d_m\)) & 512\\
  Feedforward Dimension (\(d_f\)) & 1024 \\
  Number of latent space predictions (\(c\)) & 1\\
  Total number of parameters & \(1.90 \times 10^{7} \)\\
  \hline
\end{tabular}
\end{center}
  \caption{Hyperparameters of the model used to obtain the results shown in Figure~\ref{fig:144-results}.
  Model weights are stored in single precision.}
  \label{tab:2}
\end{table}

\begin{table}[ht]
\begin{center}
\begin{tabular}{ |c|c|c|c|c|c|c| }
  \hline
  Training Stage & Batch Size & Learning Rate & \(N_R\) & \(N_H\) & Number of Epochs & Reset Optimizer\\
  \hline
  \(1\) & \(512\) & \(10^{-4}\) & \(6\) & \(0\) & \(2000\) & Yes \\
  \(2\) & \(512\) & \(10^{-4}\) & \(6\) & \(1\) & \(2000\) & Yes \\
  \(3\) & \(512\) & \(10^{-4}\) & \(6\) & \(2\) & \(2000\) & Yes \\
  \(4\) & \(512\) & \(10^{-4}\) & \(6\) &\(3\) & \(2000\) & Yes \\
  \(5\) & \(512\) & \(10^{-4}\) & \(6\) &\(4\) & \(2000\) & Yes \\
  \(6\) & \(512\) & \(10^{-4}\) & \(6\) &\(5\) & \(2000\) & Yes \\
  \(7\) & \(512\) & \(10^{-4}\) & \(6\) &\(6\) & \(3000\) & Yes \\
  {\bf8} & \(\mathbf{512}\) & \(\mathbf{10^{-4}}\) & \(\mathbf{6}\) &\(\mathbf{6}\) & \(\mathbf{4000}\) & \textbf{No} \\
  \(9\) & \(512\) & \(10^{-5}\) & \(9\) &\(7\) & \(2000\) & Yes \\
  \(10\) & \(256\) & \(10^{-5}\) & \(9\) &\(8\) & \(2000\) & Yes \\
  \(\mathbf{11}\) & \(\mathbf{512}\) & \(\mathbf{10^{-6}}\) & \(\mathbf{9}\) &\(\mathbf{9}\) & \(\mathbf{2000}\) & \textbf{Yes} \\
  \(12\) & \(512\) & \(10^{-5}\) & \(12\) &\(9\) & \(300\) & Yes \\
  \(13\) & \(512\) & \(10^{-5}\) & \(12\) &\(10\) & \(300\) & Yes \\
  \(14\) & \(512\) & \(10^{-5}\) & \(12\) &\(11\) & \(300\) & Yes \\
  \(\mathbf{16}\) & \(\mathbf{512}\) & \(\mathbf{10^{-6}}\) & \(\mathbf{12}\) &\(\mathbf{12}\) & \(\mathbf{300}\) & \textbf{Yes} \\
  \(17\) & \(256\) & \(10^{-5}\) & \(18\) &\(12\) & \(300\) & Yes \\
  \(18\) & \(256\) & \(10^{-6}\) & \(18\) &\(13\) & \(300\) & Yes \\
  \(19\) & \(256\) & \(10^{-6}\) & \(18\) &\(15\) & \(300\) & Yes \\
  \(\mathbf{20}\) & \(\mathbf{256}\) & \(\mathbf{10^{-6}}\) & \(\mathbf{18}\) &\(\mathbf{18}\) & \(\mathbf{2000}\) & \textbf{Yes} \\
  \hline
\end{tabular}
\end{center}
  \caption{Training hyperparameters of the model used to obtain the results shown in Figures~\ref{fig:72-lers-times} and \ref{fig:72-12-6-multi}. 
  Models obtained after the stages that are bolded were to generate the data in the previously mentioned figures. 
  \(N_R\) represents the number of rounds of noisy syndrome measurement in the training data, and \(N_H\) represents the number of rounds of latent space predictions.
  Each stage of training used detector outcomes obtained from simulations performed at a physical error rate of \(p=0.6\%\).}
  \label{tab:3}
\end{table}

\begin{table}[ht]
\begin{center}
\begin{tabular}{ |c|c|c|c|c|c|c|c|c| }
\hline
Training Stage & Batch Size & Learning Rate & \(N_R\) & \(N_H\) & \(p\) & c & Num. Epochs & Reset Optimizer \\
\hline
1 & \(512\) & \(10^{-4}\) & \(6\) & \(0\) & \(0.6\%\) & n/a & \(2000\) & Yes\\
2 & \(512\) & \(10^{-4}\) & \(6\) & \(0\) & \(0.6\%\) & n/a & \(1000\) & No\\
3 & \(512\) & \(10^{-4}\) & \(6\) & \(0\) & \(0.6\%\) & n/a & \(2000\) & No\\
4 & \(512\) & \(10^{-4}\) & \(6\) & \(1\) & \(0.6\%\) & 2 & \(2000\) & Yes\\
5 & \(512\) & \(5 \times 10^{-5}\) & \(6\) & \(2\) & \(0.6\%\) & 2 & \(2000\) & Yes\\
6 & \(512\) & \(2 \times 10^{-5}\) & \(6\) & \(4\) & \(0.6\%\) & 2 & \(500\) & Yes\\
7 & \(512\) & \(10^{-6}\) & \(6\) & \(5\) & \(0.6\%\) & 2 & \(500\) & Yes\\
8 & \(512\) & \(10^{-5}\) & \(6\) & \(6\) & \(0.6\%\) & 2 & \(500\) & Yes\\
9 & \(512\) & \(10^{-5}\) & \(6\) & \(6\) & \(0.6\%\) & 2 & \(1000\) & No\\
10 & \(512\) & \(10^{-6}\) & \(6\) & \(6\) & \(0.6\%\) & 2 & \(2000\) & Yes\\
11 & \(256\) & \(10^{-5}\) & \(6\) & \(6\) & \(0.6\%\) & 1 & \(2000\) & Yes\\
12 & \(256\) & \(10^{-6}\) & \(6\) & \(6\) & \(0.4\%\)& 1 & \(2000\) & Yes\\
13 & \(256\) & \(5 \times 10^{-6}\) & \(12\) & \(6\) & \(0.6\%\)& 1 & \(500\) & Yes\\
14 & \(256\) & \(5 \times 10^{-6}\) & \(12\) & \(7\) & \(0.6\%\)& 1 & \(500\) & Yes\\
15 & \(256\) & \(2 \times 10^{-6}\) & \(12\) & \(8\) & \(0.6\%\)& 1 & \(500\) & Yes\\
16 & \(256\) & \(10^{-6}\) & \(12\) & \(12\) & \(0.6\%\)& 1 & \(2000\) & Yes\\
17 & \(256\) & \(10^{-6}\) & \(12\) & \(12\) & \(0.4\%\)& 1 & \(2000\) & Yes \\
\hline
\end{tabular}
\end{center}
  \caption{Training hyperparameters of the model used to obtain the results shown in Figure~\ref{fig:144-results}. 
    \(N_R\) represents the number of rounds of noisy syndrome measurement in the training data, and \(N_H\) represents the number of rounds of latent space predictions. 
    We additionally note that in training stage 13 we used a learning rate scheduler, which increased the learning rate linearly from zero over the course of \(1000\) batches, and then adjusted the learning rate as \(10^{-5} \cdot 1/n^{0.0625}\) where \(n = \text{num steps} - 1000\).}
    \label{tab:4}
\end{table}
\end{document}